# Intense Geomagnetic Storms during Solar Cycles 23-25

*Gopalswamy N.[1], Akiyama S.[1,2], Yashiro S.[1,2], Makela P.[1,2], Xie H.[1,2]*

[1]Heliophysics, NASA GSFC, Greenbelt, MD, USA; nat.gopalswamy@nasa.gov
[2]The Catholic University of America, Washington DC

## 1. Abstract.

Intense geomagnetic storms are characterized by a minimum value of the Dst index at or below –100 nT. It is well known that these storms are caused by the southward magnetic fields in coronal mass ejections (CMEs) and corotating interaction regions (CIRs). While CIR storms are confined to Dst values at or above –150 nT, CME storms can reach Dst –500 nT or lower. In this report, we illustrate the need to understand the storm evolution based on solar source and solar wind parameters using a recent storm (2023 April 24) by way of providing the motivation to catalog such events for a better understanding of the main phase time structure of geomagnetic storms.

## 2. Introduction

Intense geomagnetic storms (Dst ≤ -100 nT) are known to have serious space weather consequences. They result in energization of particles in Van Allen belts, particle precipitation in the polar atmosphere, atmospheric heating, and geomagnetically induced currents (GICs; see Temmer 2021, and references therein). The time structure of geomagnetic storms represented by an index such as Dst can introduce large variability in the strength, duration, and consequences of geomagnetic storms. Rapid time structures can result in large GICs (Pulkkinen et al. 2005; Kappenman et al. 2005; Belakhovsky et al. 2019). One type of time structure of a geomagnetic storm is multiple steps in the main phase. The multistep storms constitute about two-thirds of all intense geomagnetic storms (Zhang et al. 2008). The classic case of two Bz<0 intervals in the sheath and MC separated by a Bz>0 interval results in the double-dip storm. Most multi-dip storms are due to coronal mass ejections (CMEs), but ~ 11% of dips are found in storms caused by corotating interaction regions (CIRs). Another source of two-dip storms is the passage of an interplanetary shock through the body of a preceding CME resulting in an abrupt enhancement of the Bz magnitude (Richardson and Zhang, 2008; Gopalswamy et al. 2015). Multiple dips can also occur in complex ejecta formed by two or more CMEs and their sheaths.

Another type of storm main phase time structure was reported in Gopalswamy et al. (2022). The time structure appears as a sudden change in the slope of the Dst profile in response to a change in the solar wind dynamic pressure (Gopalswamy et al. 2022). The slope change coincided with the density increase inside the second half of the 2018 August 25 MC. Simulation of the storm using the Comprehensive Inner Magnetosphere-Ionosphere (CIMI) model (Fok et al., 2014) confirmed that the Dst slope change coincided with an increase in the ring current energy. Murayama (1982) introduced the effect of the solar wind dynamic pressure (Pf) via the ring current injection (Q) in the form Q ~ VBz×Pf1/3. Following the work of several authors, Wang et al. (2003) refined Q and the ring-current decay rate ($\tau$) as functions of solar wind electric field Ey = VBz and Pf. Xie et al. (2008) used Wang et al. formulas and showed that the Dst peak value of a storm increases when there is a large enhancement in Pf during the storm main phase. Le et al. (2020) and Zhao et al. (2021) found that the SYM-H index is highly correlated with the time-integral of the injection over the main phase. Thus, one needs to derive Q and τ from solar wind parameters for a better understanding of the time structure of the main phase.

The origin of the main phase time structure is also related to the solar source properties such as the tilt angle and axial-field direction of the CME flux rope, and the flux rope kinematics as influenced by the coronal and interplanetary environment. These parameters can be derived from the Flux Rope from Eruption Data Technique (Gopalswamy et al. 2018). In this report, we examine the largest geomagnetic storm over the first three years of solar cycle 25 (2019 December 1 to 2023 April 30). We also compare the occurrence rate and intensity of geomagnetic storms over the corresponding phases in solar cycles 23 and 24.

### 3. Observations, Analysis, and Results

As of this writing the largest geomagnetic storm (Dst = –212 nT) in solar cycle (SC) 25 occurred on 2023 April 24 around 06:00 UT as obtained from the World Data Center in Kyoto (Nose et al. 2015). This is a double-dip storm with both dips being below –100 nT. The solar wind parameters showing the associated ICME were obtained from the OMNI database. The CME associated with the storm was observed by the Large Angle and Spectrometric Coronagraph (LASCO, Brueckner et al. 1995) on board the SOHO mission. The CME was also observed by the coronagraphs in the Sun Earth Connection Coronal and Heliospheric Investigation (SECCHI, Howard et al. 2008) instrument suite on board the STEREO mission. Information on the source active region was obtained from the Atmospheric Imaging Assembly (AIA, Lemen et al. 2012) on board the SDO mission. Plots and movies made available at the CDAW data center (https://cdaw.gsfc.nasa.gov) were also used for this investigation.

#### *3.1. Solar Wind Parameters and the Geomagnetic Storm*

The solar wind parameters associated with the storm are shown in Figure 1. We see that the interplanetary CME (ICME) in this event is a magnetic cloud (MC) according to the definition of Burlaga et al. (1981): enhanced magnetic field, smooth rotation of one of the magnetic field components and low proton temperature or plasma beta. The shock-driving MC was also expanding with the front and back speeds of 620 km/s and 490 km/s, respectively. The magnetic field in the MC was significantly enhanced, with a peak value of ~35 nT. The Bz component rotated from south to north while the By component remained east-west indicating that this is a low-inclination MC. The shock sheath had enhanced density and magnetic field while these two parameters were significantly depressed inside the MC. The shock seems to be propagating into a small preceding flux rope whose tail end was compressed by the shock resulting in a large negative Bz. The MC on April 24 is bipolar with south-north configuration, typical of an odd-numbered solar cycle. The ambient solar wind ahead of the MC has a speed of ~350 km/s as can be seen at the beginning of the plot (before 4:00 UT on April 23). At the shock arrival the IMF Bz suddenly decreased to -20 nT while the density increased to about 20 cm$^{-3}$. The density was high throughout the sheath region, dropping to below 5 cm$^{-3}$ in the MC, remaining low throughout the MC interval.

The Dst index was already decreasing around 10:00 UT on 2023 April 23 from –13 nT to –49 nT at 16 UT ahead of the shock. When the shock arrived, the Dst slope changed suddenly and the Dst dropped to –56 nT. From then on the Dst index dropped rapidly reaching –165 nT (the first dip due to Bz reaching –10 nT in the sheath). The Dst slope changed from 6 nT/hr to ~22 nT/hr. Bz turned positive around 21 UT on April 23, so the storm started recovering for the next 3 hrs until the arrival of the MC. The Bz component turned negative in the beginning of the MC reaching –30 nT right after the front boundary of the MC. Accordingly, Ey reached –20 mV/m, which is twice the value during the Bz <0 interval in the sheath (see Fig. 1).

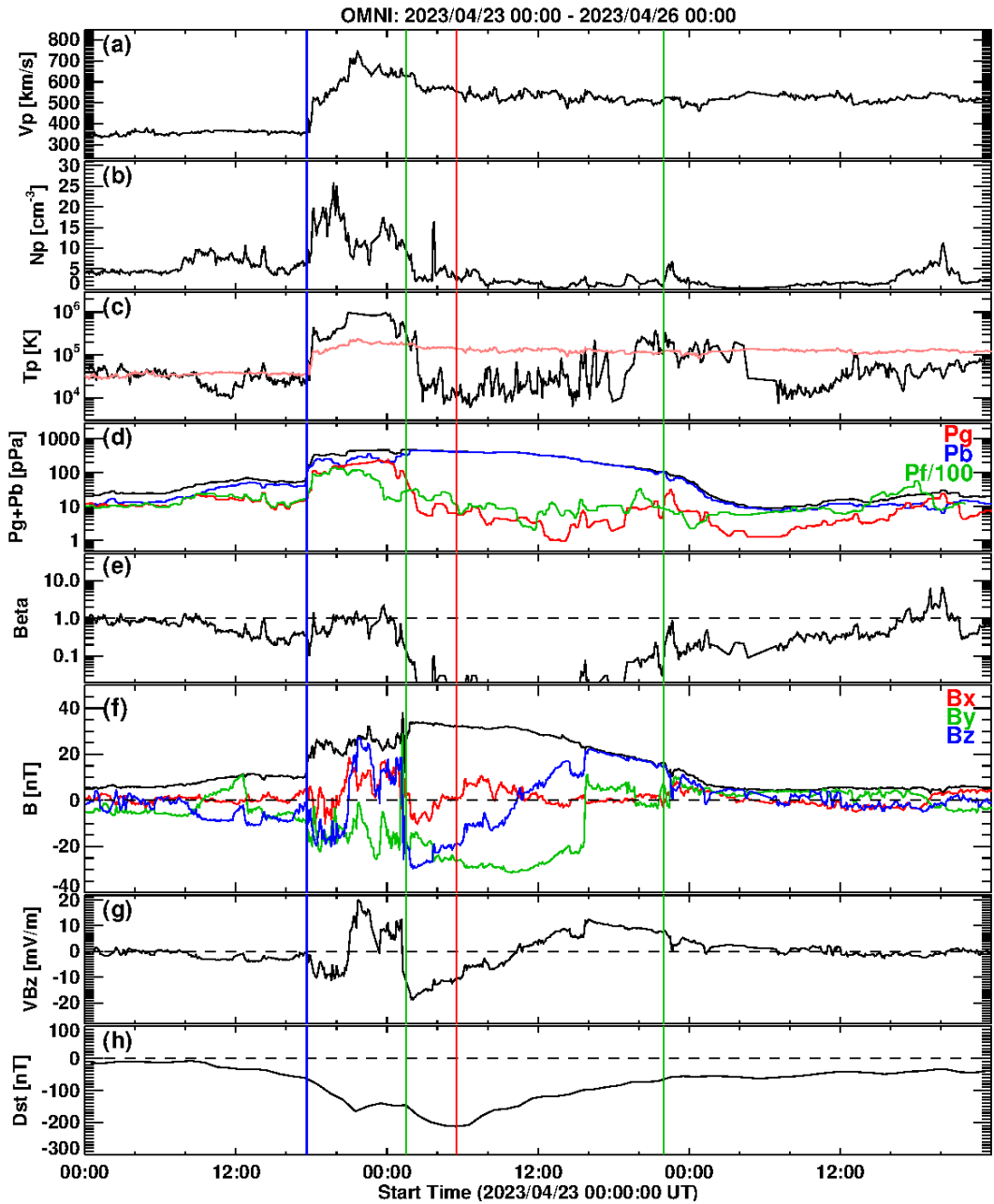

*Figure 1. Solar wind observations from OMNI for the period 2023 April 23 – 26. (a) Solar wind speed (Vp), (b) proton density (Np), (c) proton temperature (Tp) along with the expected temperature (orange line), (d) gas (Pg – red curve), magnetic (Pb – blue curve), and flow (Pf – green curve) pressures and the total pressure (Pg+Pb – black curve), (e) plasma beta, (f) total magnetic field strength (B) along with the three components Bx (red curve), By (green curve), and Bz (blue curve) in GSE coordinates, (g) solar wind electric field (solar wind speed times the Bz component of the magnetic field), (h) the Dst index showing the intense double-dip geomagnetic storm. The vertical red line marks the time of minimum Dst (–212 nT at 06:00 UT on April 24). The first dip (–165 nT) in Dst occurred at 22:00 UT on April 23. The Dst data are from the World Data Center, Kyoto. The vertical blue line marks the shock (17:38 UT on April 23). The vertical green lines mark the boundaries of the magnetic cloud based on Tp and plasma beta (beginning and end of MC interval). B is also enhanced significantly during the MC interval (1:30 UT to 22:00 UT on April 24).*

The larger difference in Ey is due to the higher solar wind speed during the Bz <0 interval in the MC. In spite of the large Ey, the Dst index displayed a shallower slope, taking ~5 hrs to decrease by 65 nT (13 nT/hr). The reason behind the shallow slope is the low density in the MC (<5 cm-3). The effect is opposite to that of the steep slope caused by high density (or high dynamic pressure) as shown in Gopalswamy et al. (2022) for the 2018 August 26 storm. After the minimum value, the Dst increases mainly due to the weakening of Ey until it changes sign when Bz turns positive. Further recovery is slow mainly because of the low density in the MC, consistent with the dependence of ring current decay under low dynamic pressure (Wang et al. 2003). Thus, for a full understanding of the Dst index (ring current evolution), one needs to consider the main-phase time structure caused by Ey and dynamic pressure.

### *3.2. Events at the Sun and the Source Active Region*

The solar counterpart of the MC and the source region on the Sun can be determined by examining CMEs that occurred 1-4 days before the storm. This can be done using the composite plots involving the Dst index, CME height-time plot, and GOES soft-ray light curve available at https://cdaw.gsfc.nasa.gov/CME_list/daily_plots/dsthtx/2023_04/dsthtx_20230421.html (see Fig. 2). We see that was a series of 5 CMEs from the east limb On April 19. The CMEs on April 20 and the first one on April 21 are also limb CMEs, so none of these CMEs could have caused the intense geomagnetic storm. Therefore, it is easy to identify the CME that caused the geomagnetic storm as the CME on April 21 appearing in the SOHO/LASCO FOV at 18:12 UT. The CME was associated with a major flare (GOES intensity M1.7) originating from S22W11 on the solar disk.

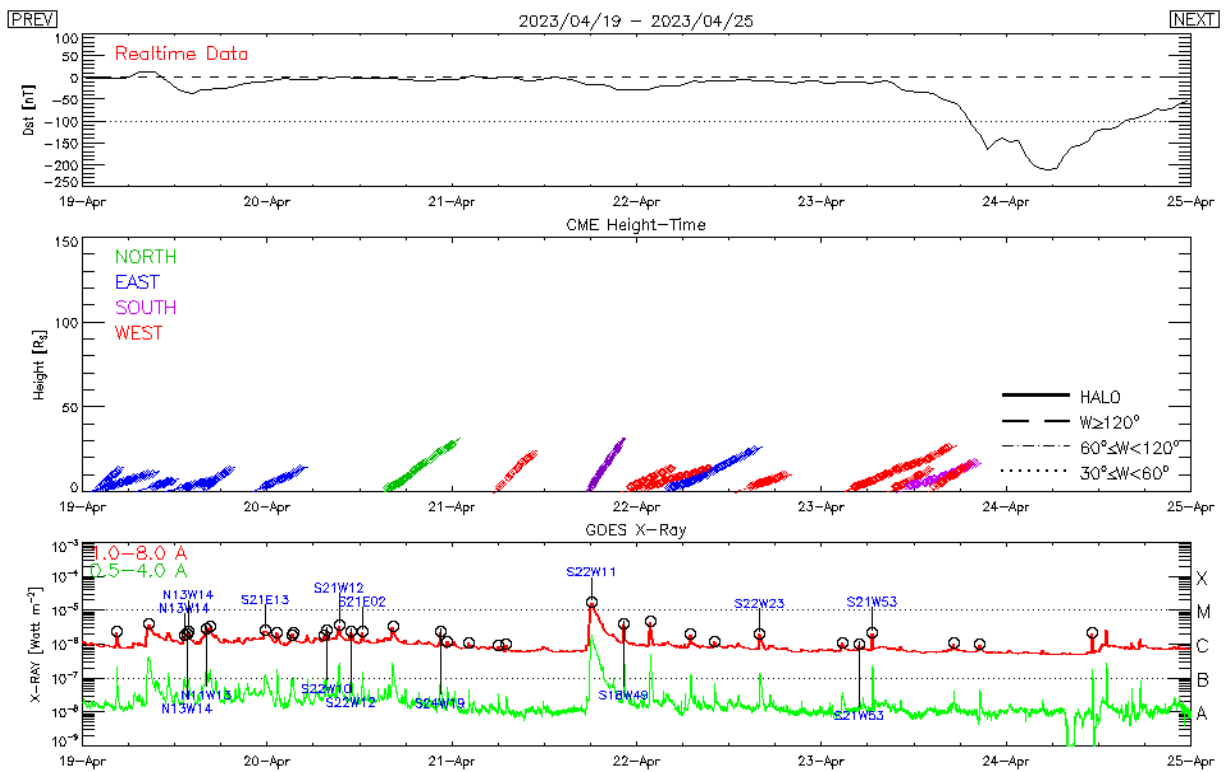

*Figure 2. Overview plot showing the Dst index over a six-day period starting from April 19, 2023. The CME that caused the geomagnetic storm originated from AR 13283 located at heliographic coordinates S22W11.*

Figure 3 shows the early manifestation of the CME using an EUV difference image at 17:56 UT obtained by SDO/AIA at 192 A. The EUV disturbance is from the southwest quadrant as expected from the flare location (S22W11). The figure also shows the fully formed halo CME with the ejecta part (bright feature) heading in the southwest direction. The diffuse features are thought to be manifestation of the shock sheath. The STEREO Ahead spacecraft located at E10 observed the CME with a similar morphology. The CME had a sky plane speed of ~1284 km/s, slightly higher than the average speed of CMEs causing intense geomagnetic storms. The CME produced an intense type II radio burst and a minor solar energetic particle (SEP) event indicating that the CME was driving a shock from near the Sun. The CME showed a deceleration of $-28$ m s$^{-2}$, as expected for a fast CME. The MC speed measured at 1-au was ~600 km/s, consistent with the CME deceleration in the coronagraph FOV.

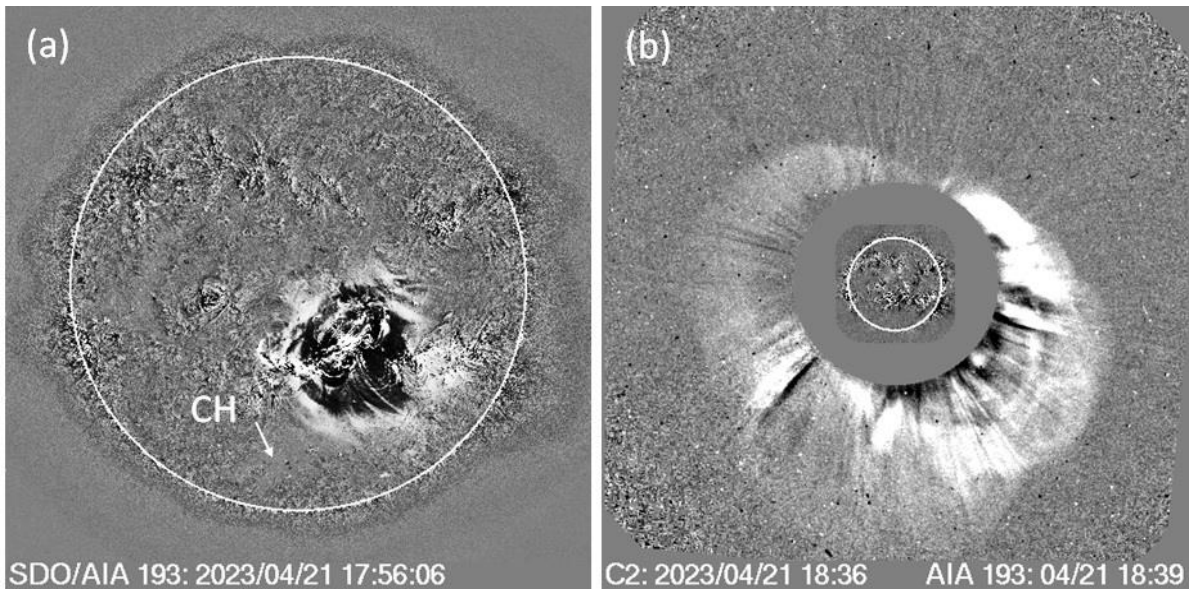

***Figure 3.*** *Figure 3 (a) The solar source SDO/AIA at 193 Å difference image and (b) the CME as viewed in the SOHO/LASCO FOV. The active region 13283 was located at S22W11. The B0 angle was +6⁰. The larger latitudinal distance was likely reduced by deflection of the CME by the coronal hole located to the south of the eruption region. The deflection might have pushed the CME trajectory to the north and west of the Sun-Earth line.*

Figure 4 shows the CME source region (NOAA AR 13283) in the longitudinal magnetogram obtained by SDO's Helioseismic Imager (HMI) and the post eruption arcade (PEA) imaged by SDO's AIA at 193 Å. The coronal hole to the southeast of the active region has negative polarity. The main neutral line in the active region is in the east west direction. The PEA is also extended in the east west direction with a slight tilt. The neutral line orientation matches with the low inclination MC inferred from in-situ observations (see Fig. 1). The source location is close to the disk center, so the CME is likely hit Earth as confirmed by in-situ observations.

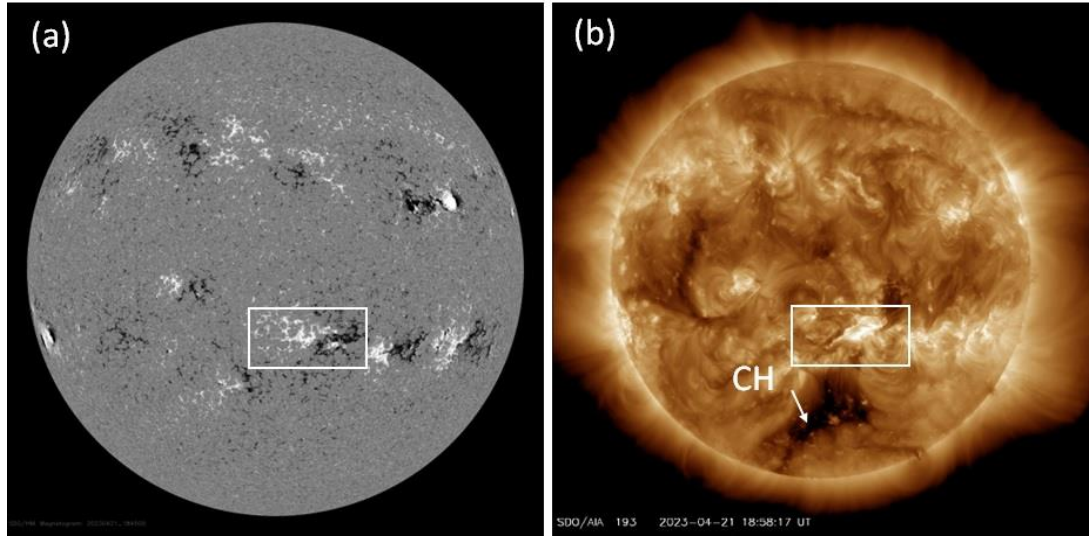

*Figure 4.* (a) SDO/HMI magnetogram with the source active region marked by the white rectangle. (b) SDO/AIA image showing the post eruption arcade. The coronal hole (CH) to the southeast of the eruption region is marked by an arrow.

### 3.3. Shock arrival time at 1 au

In order to compare the observed shock travel time with that expected from empirical models, we need to get the true speed of the CME in the earthward direction. STEREO was observing from very close to the Sun-Earth line, so we cannot use forward modeling with multiview observations. Instead, we use an empirical model obtained by Gopalswamy et al. (2015b) developed by comparing cone model results with direct measurements for CMEs in quadrature. According to this model, the space speed ($V_{3D}$) of the CME can be estimated from the sky-plane speed ($V_{sky}$) using the relation, $V_{3D} = 1.1\ V_{sky} + 156$ km/s as 1568 km/s. Based on the source location (S22W11) and the measured Vsky = 1284 km/s, we estimate the earthward speed (u) as 1428 km/s. Using u in the empirical shock arrival model for the travel time T as $T = AB^u + C$ (with A = 151.002, B = 0.998625, and C = 11.5981, see Gopalswamy et al. 2005). The result is T = 32.8 hrs. The actual travel time measured from the time of first appearance of the CME (08:12 UT on April 21) to the shock arrival at 1 au (17:38 UT on April 23) is ~47 hrs, indicating a major discrepancy. This discrepancy is likely due to a smaller earthward speed than computed above because of the CME deflection to the north and west by the coronal hole. Such deflections are very common on the Sun (Gopalswamy et al. 2009). If the CME is deflected by an angle of ~30⁰ to the west, then the earthward speed becomes u = 1097 km/s, yielding a travel time of ~45 hrs, which is close to the observed travel time.

*Table 1.* Number of intense storms during the first 41 months of solar cycles 23-25.

| Solar Cycle | #CME storms | #CIR Storms | Total |
|---|---|---|---|
| SC 23 | 18 | 3 | 21 (18)[1] |
| SC 24 | 5 | 0 | 5 (0) |
| SC25 | 4 | 0 | 4 (1) |

[1]The numbers in the parentheses of this column correspond to the rise phase.

## 4. Discussion

We studied the solar and interplanetary causes of the 2023 April 24 geomagnetic storm, which is the largest in the first 41 months in solar cycle 25. It must be noted that the Sun is already in the maximum phase starting in late 2022. Only one intense storm occurred during the rise phase of SC 25 on 2021 November 4 with Dst = –105 nT. In contrast, SC 23 witnessed 18 intense storms in its rise phase including two storms with intensity <–200 nT. On the other hand, SC 24 had no intense storm in its rise phase. Table 1 compares the number of intense storms in the first 41 months in each. Note that this interval includes part of the maximum phase of the cycles. The numbers in the last column within parentheses correspond to the rise phase in each cycle. These numbers are consistent with the finding that the strength of SC 25 is likely to be similar or slightly greater than that of SC 24 (Gopalswamy et al. 2023).

How does the CME speed compare with the average speed of CMEs that result in major storms? Figure 5 shows the distribution of sky-plane speeds of 91 CMEs observed during SCs 23, 24, and part of SC 25 that caused intense storms. Also shown are the halo fraction and the average width of non-halo CMEs. We see that the intense storms are caused by fast and wide CMEs, most of them being halo CMEs. The average speed of the CMEs is 950 km/s, more than a factor 2 faster than the general population of CMEs. Note that these are sky-plane speeds. The true speeds are likely to be higher when the sky-plane speeds are deprojected. The speed of the 2023 April 24 storm is shown on the speed histogram, indicating that it has above-average speed. The CME in hand falls in the halo CME bin shown in Fig. 5(b). The location of the CME is also consistent with the distribution of source positions as indicated by the arrow mark in Fig. 5 (c).

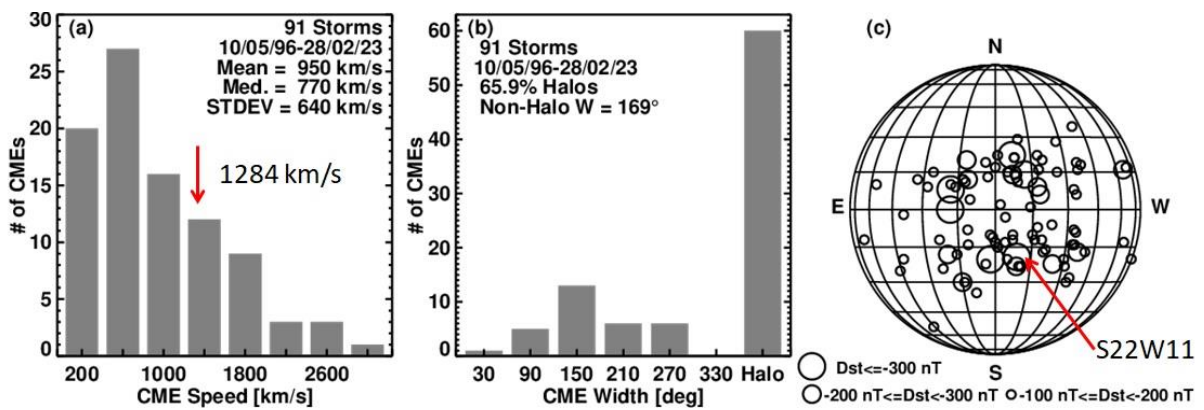

*Figure 5*. Distributions of speed (a), width (b), and source locations (c) of CMEs responsible for intense geomagnetic storms. The mean and median speeds are along with standard deviations. The red arrows point to the parameters of the 2023 April 24 storm.

## 5. Conclusions

The main finding of this work is that we need to consider both the solar wind electric field and the dynamic pressure. The slope changes in the main phase and the early decay phase are marked by the changes in proton density in the sheath and MC. When the dynamic pressure is low, say below 3 nPa, the Dst profile is controlled by the electric field. When the dynamic pressure is high, we need to consider the contributions from both electric field and the dynamic pressure. On the solar side we need to take into account of the presence of a coronal hole near the eruption region. The effect of the coronal hole is to deflect the CME in question away from the Sun-Earth line in the longitudinal direction. Note that depending on the relative location of the coronal hole and the eruption region, some CMEs can be deflected toward the Sun-Earth line. Motivated by the results of this investigation, we are developing a catalog of intense storms linking them to the interplanetary event described in Fig. 1 and the solar sources and CMEs as illustrated in Figs. 3 and 4. Similar information on solar and solar wind links to CIR storms will also be included in the catalog (Gopalswamy et al. 2023, under preparation).